\documentclass[twocolumn,preprintnumbers,amsmath,amssymb]{revtex4}
\usepackage[latin1]{inputenc}
\usepackage{amsmath}
\usepackage{amssymb}
\usepackage{amscd}
\usepackage{graphicx}
\usepackage{subfigure}
\usepackage{graphicx}

\begin{document}

\title{Timescale of entropic segregation of flexible polymers in confinement: implications for chromosome segregation in filamentous bacteria}

\author{Axel Arnold}
\email{arnold@amolf.nl}
\affiliation{FOM-Institute AMOLF, Kruislaan 407, 1098 SJ Amsterdam, The Netherlands}

\author{Suckjoon Jun}
\affiliation{Facult\'{e} de M\'{e}decine, INSERM Site Necker, U571, 156 rue de Vaugirard, 75015 Paris, France}
 \altaffiliation{Present address: FAS Center for Systems Biology, Harvard University, Cambridge, MA 02138, USA}
\date{\today}

\begin{abstract}
  We report molecular dynamics simulations of the segregation of two overlapping
  chains in cylindrical confinement. We find that the entropic repulsion between
  the chains can be sufficiently strong to cause segregation on a time scale
  that is short compared to the one for diffusion.  This result implies that
  entropic driving forces are sufficiently strong to cause rapid bacterial
  chromosome segregation.
\end{abstract}
\preprint{entropic segregation}
\maketitle


\section{Introduction}

Confined polymers play an important role in many industrial
processes and  biological systems. Examples range from membrane
filtration and oil recovery to gel electrophoresis and protein
translocation~\cite{Kasianowicz02, Duke04, Alberts}. Importantly,
recent technological development in nano-/micro-fluidics has made
it possible to manipulate and trap biomolecules such as
double-stranded (ds) DNA in  confined environments with a
characteristic lengthscale that is much smaller than the radius of
gyration of the polymers~\cite{Austin04, Craighead06, Squires05}.
Also under biological conditions, DNA is often strongly confined,
e.g. packed into a viral capsid~\cite{Gelbart01},
bacteria~\cite{Jun06} or the eukaryotic cell
nucleus~\cite{Cremer01}.

In this article, we report molecular dynamics simulations that
allow us to determine the typical speed of the segregation of
initially mixed polymers in cylindrical confinement. This problem
has particular relevance for the understanding of chromosome
segregation in bacteria, where the nature of its underlying
mechanism is currently under debate. Here, the basic issue is
whether the major driving force for segregation of duplicating
chromosomes in strong confinement is physical (driven by entropy
or mechanical ``pushing'')~\cite{BatesKleckner, Jun06} or
biological (such as cytoskeletal and motor proteins)~\cite{Gerdes,
Gitai}.

Our results show that the effective repulsion between two chains in a
cylindrical geometry of confinement can be very strong.  Typically, the
segregation requires a time proportional to $N^2$, which is much faster than the
$N^3$ timescale of chain diffusion, where $N$ is the chain length.
This suggests that for filamentous
bacteria such as \emph{Streptomyces coelicolor}~\cite{Hopwood} or cyanobacterium
\emph{Anabaena}~\cite{Zhao}, the main driving force of chromosome segregation
might be entropic and any additional mechanisms are for ``optimization.'' As we
shall discuss later, our proposal is fully consistent with the recent results
that chromosome segregation in some filamentous bacteria is a \emph{random}
process~\cite{Zhao}.

\section{Theory}

\begin{figure}[tp]
  \centering
  \includegraphics[width=.48\textwidth]{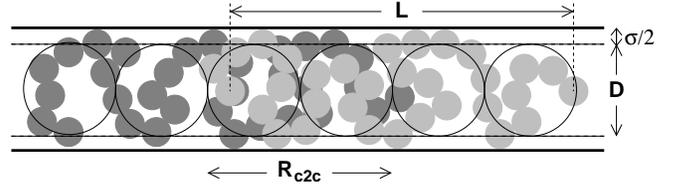}
  \caption{Two partly overlapping chains in a spherical cylinder of width $D$.
    The chains consist of $N$ beads of size $\sigma$ each.  $L$ denotes the
    chain extension, and $R_\text{c2c}$ the distance of the centers of mass. }
  \label{fig:blobs}
\end{figure}

Consider two linear chains with excluded-volume interactions, which are
initially intermingled and confined in an infinitely long cylinder with a
diameter $D$ that is much smaller than the radius of gyration $R_g$ of the
unconfined chains (Fig.~\ref{fig:blobs}). As the two chains can gain
conformational entropy by demixing, they effectively repel each other.  Note
that the free energy cost for simultaneous overlap of $n$ chains scales as
$n^{9/4}$ in the dilute regime and then increases faster as $n^3$ in the more
concentrated regime, independent of the chain length~\cite{Grosberg, Jun07}.
Thus, for two intermingling chains illustrated in Fig.~\ref{fig:blobs}, each
blob-blob overlap contributes $\sim k_BT$ ($n=2$) to the free-energy cost.  The
potential of mean force for segregation between the two chains then is
proportional to the total number of overlapping blobs, {\i.e.}, $\beta
\mathcal{F}(R_\text{c2c}) \simeq (L_\text{eq}-R_\text{c2c})/D$, where
$L_\text{eq}$ is the equilibrium length of an isolated individual chain in the
pore, and $R_\text{c2c}$ the center-to-center distance between the two chains.
The effective repulsive force is then obtained as
\begin{equation}
\label{eq:feffective}
F_\text{eff} = -\frac{\partial \mathcal{F}}{\partial R_\text{c2c}} = \frac{k_BT}{D},
\end{equation}
\noindent and, thus, the equation of motion for the center of mass is
\begin{equation}
\label{eq:eqmotion}
M \dot{V}_\text{c2c} = \frac{k_BT}{D} - \Gamma V_\text{c2c}.
\end{equation}
where $M=N m$ and $\Gamma$ are the total mass and the effective friction of the
chain, respectively ($m$ is the mass of a single monomer). Ignoring
hydrodynamic interactions between monomers, one can assume that the frictions
$\gamma$ on the individual monomers are additive, i.e., also $\Gamma= N \gamma$.
Then, the solution of Eq.~\ref{eq:eqmotion} with an initial condition
$V_\text{c2c}=0$ at $t=0$ can be obtained as
\begin{equation}
  \label{eq:sol1}
  V_\text{c2c}(t) = \frac{k_BT}{\gamma D N} (1 - e^{-\frac{t}{\tau^*}}),
\end{equation}
where $\tau^* = m/\gamma$ is the ``inertial'' timescale. In
practice, $t\gg \tau^*$ and hence the characteristic segregation
speed is constant and given by
\begin{equation}
  \label{eq:sseg}
  V_{c2c} \sim \frac{k_BT}{\gamma DN}.
\end{equation}

The equilibrium length of confined chains, $L_\text{eq}$, is
proportional to $N$. Therefore, the time for reaching complete
segregation, $R_\text{c2c}=L_\text{eq}$, scales as
\begin{equation}
  \label{eq:tseg}
  t_s \sim L_\text{eq}/V_\text{c2c} \sim N^2.
\end{equation}
This is time is much shorter than $t_\text{diff}$, the typical
time it takes a single chain to diffuse over a distance equal to
its own length:
\begin{equation}
  \label{eq:trept}
  t_\text{diff}\sim \frac{L_\text{eq}^2}{2 D_\text{diff}} \sim N^3.
\end{equation}
However, the above considerations do apply for the initial
situation of complete overlap, $R_{c2c}\approx 0$. In this case
the system is in a state of unstable equilibrium, since the
effective segregation force is $F_\text{eff}\approx 0$. Hence, the
system will initially show purely diffusive behaviour until a
certain separation, typically $R_{c2c}\approx D$, is reached. We
refer to the time until segregation sets in as the ``induction
time'' $t_i$ that should scale as $N^3$. With increasing $D$,
diffusion becomes easier, because the monomer concentration
decreases, and $t_i$ decreases, while $t_s$ increases with $D$.
Below, we show that for all practically relevant diameters, the
segregation process is rate limiting.

For real bacteria, entropic segregation already sets in during replication.
Therefore, segregation always takes place for comparatively short pieces of DNA,
so that the induction time does not play a role.

\section{Simulation method}

In the molecular dynamics simulations, we model the polymers using a
bead--spring model in a cylindrical compartment of diameter $D$; each chain
consists of $N$ beads of diameter $\sigma$. The bead--bead and bead--compartment
interactions were modeled by a Weeks-Chandler-Andersen potential
(WCA)~\cite{wca}, which corresponds to the repulsive part of the Lennard-Jones
potential:
\begin{equation}
  U_\text{WCA}(r)=\epsilon_\text{WCA}\left[\left(\frac{\sigma}{r}\right)^{12} -
    \left(\frac{\sigma}{r}\right)^6 + \frac{1}{4}\right]
  \label{eq:lj}
\end{equation}
for $r<\sqrt[6]{2}\sigma$ and 0 elsewhere. $r$ denotes the
distance between two bead centers for the bead--bead interactions,
and the distance between the bead center and the compartment minus
$\sigma$ for the bead--compartment interactions. At $r=\sigma$,
the interaction energy is $\epsilon_\text{WCA}=1 k_BT$; since the
potential is quite steep, $r$ will typically stay above
$0.9\sigma$. This models soft beads of diameter $\sigma$, whose
centers cannot come much closer than $\sigma$ to each other, and
cannot penetrate the wall (i.~e. the wall imposes a constraint on
the sphere centers, as depicted in Fig.~\ref{fig:blobs}). In the
simulation, $\sigma$ defines the basic length scale and
$\epsilon_\text{WCA}$ the energy scale. Our unit of mass is given
by $m$, the mass of a bead.  We choose the temperature such that
$k_BT/\epsilon=1$. Having specified our basic units, the time unit
is given by $\tau_\text{WCA}=\sigma
\sqrt{m/\epsilon_\text{WCA}}=1$. In the following, we will omit
these units.

The springs between the beads in a chain were formed by the FENE (finite
extensible nonlinear elastic) potential
\begin{equation}
  U_F(r) = -\frac{1}{2} \epsilon_F r_F^2 \ln \left[ 1 - \left(
      \frac{r}{r_F} \right)^2 \right]\,,
  \label{eq:fene}
\end{equation}
where $r$ is the distance of the bead centers, $r_F$ is the radius
at which the potential becomes singular, and $\epsilon_F$ is the
interaction strength. In the present simulations, we chose
$\epsilon_F=10$ and $r_F=2$.  In combination with the WCA
potential this results in a typical bond length of $1.027$.

\begin{table}[tbp]
  \raggedright
  N=100\hfill\raisebox{1em}{
    \begin{tabular}[t]{|l|l|l|l|l|l|l|}
      \hline
      $D$      &  1.5 &  2   &  2.5 &  3   &  4   &  5 \\
      \hline
      $L_\text{eq}$ & 72.5 & 64.4 & 58.1 & 52.8 & 44.2 & 37.6\\
      \hline
    \end{tabular}}
  \vspace{.5em}

  N=200\hfill\raisebox{1em}{
    \begin{tabular}[t]{|l|l|l|l|l|l|l|l|l|}
      \hline
      $D$      &   1.5 &   2   &   2.5 &   3   &  4   &  5   &  7   & 9 \\
      \hline
      $L_\text{eq}$ & 146.0 & 130.3 & 118.0 & 107.6 & 91.2 & 78.9 & 61.1 & 49.4\\
      \hline
    \end{tabular}}
  \vspace{.5em}

  N=300\vspace{.5em}\\
  \begin{tabular}{|l|l|l|l|l|l|l|l|l|l|l|}
    \hline
    $D$      &   2   &   3   &   4   &   5   &   6   &  7   &  8   & 9   &%
    11 & 13\\
    \hline
    $L_\text{eq}$ & 190.4 & 162.1 & 133.2 & 121.2 & 105.2 & 94.6 & 88.2 &%
    77.6 & 64.8 & 55.8\\
    \hline
  \end{tabular}

  \vspace{.5em}
  \begin{tabular}[t]{|l|l|l|l|}
    \hline
    N & $T_{\text{warm}}$ & $T_{\text{config}}$ & $N_{\text{config}}$ \\
    \hline
    100 & $10^5$           &  2000 &  800 \\
    200 & $8\cdot 10^5$    &  8000 & 1000 \\
    \hline
  \end{tabular}
  \begin{tabular}[t]{|l|l|l|l|}
    \hline
    N & $T_{\text{warm}}$ & $T_{\text{config}}$ & $N_{\text{config}}$ \\
    \hline
    300 & $1.8\cdot 10^6$  & 36000 &  200\\
    \hline
  \end{tabular}
  \caption{The simulation parameters for the different runs. The first three
    tables give for different chain lengths $N$ the simulated pore diameters
    $D$ and the corresponding equilibrium end---to---end distances $L_\text{eq}$
    of a single chain. The last table contains the number $T_{\text{warm}}$
    of timesteps used for equilibration of the interconnected chains, the
    number of timesteps $T_{\text{config}}$ between recorded configurations,
    and the number $N_\text{config}$ of independent simulations runs with
    different random seeds.}
  \label{tab:simparams}
\end{table}

We simulate this system using the simulation package
ESPResSo~\cite{espresso}. To propagate the system, we employ a
velocity-Verlet MD integrator with a fixed time step of $0.01$;
the system is kept at constant temperature by means of a Langevin
thermostat with a fixed friction of
$\gamma=m\tau_\text{WCA}^{-1}$, so that $\tau^*=\tau_\text{WCA}=1$.
Other parameters vary for the different simulation runs, see
Table~\ref{tab:simparams}. Our simulation procedure contains four
steps:

The system is initially prepared in a ``ladder'' configuration
formed by two interconnected zig--zag strands, \i.e.,  the system
consists of two linear chains where the $i$--th bead of one chain
is bonded to the $i$--th bead of the other chain, in addition to
the bonds to its neighbors within the same chain.

To equilibrate the system, we simulate for $T_{\text{warm}}$ steps
with a ``soft'' WCA potential,  \i.e.  a WCA potential that has
been modified such that the potential is linear for distances
smaller than a radius $r_\text{fc}$. We reduce $r_\text{fc}$
gradually during the equilibration phase, so that the potential
converges to the plain WCA interaction. This procedure allows more
overlap between beads during the initial equilibration phase,
which helps the bonds to quickly relax to the equilibrium length.

After the equilibration of the interconnected chains, we remove
the interconnecting bonds to obtain two separate chains whose
centers of mass very nearly coincide. The chains are stretched by
about $10-20\%$ compared to a single chain in confinement due to
the cross linking; however, their length relaxes quickly to almost
the same length as a single chain once the cross linking is
released. The timescale for this relaxation is negligible compared
to the segregation time.

We continue to simulate the system until the two chains have
segregated, \i.e., until the chains do not overlap and their
centers of mass are separated by at least the equilibrium length
$L_\text{eq}$ of a single chain, which had been determined
beforehand by separate simulations.  During this run, we record
configurations every $T_{\text{config}}$ simulation steps.

\begin{figure}[tb]
  \centering
  \includegraphics[width=.48\textwidth]{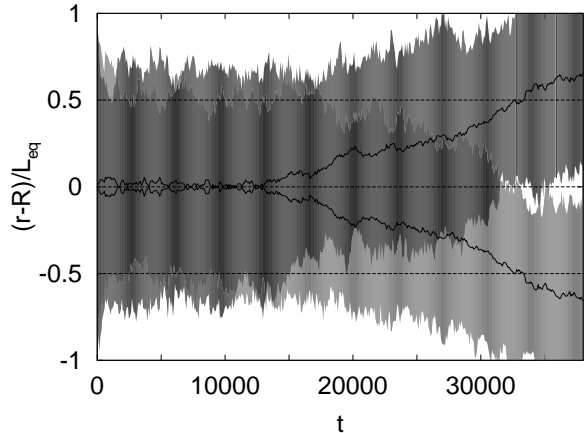}
  \caption{Example simulation run for $D=7$, $N=200$, starting from the removal
    of the interconnecting bonds. The two gray bands give the total extents
    along the tube axis of the two chains for a single run, the black lines the
    positions of their centers of masses. The positions are relative to the
    total center of mass $R$ of the system and rescaled by the equilibrium
    length $L_\text{eq}$ of a single confined chain.}
  \label{fig:segrun}
\end{figure}

This procedure is repeated $N_\text{config}$ times (see
Table~\ref{tab:simparams}), resulting in $N_\text{config}$
independent data sets similar to Fig.~\ref{fig:segrun}. For each
of these data sets, we calculate the distance $R_\text{c2c}(t)$ of
the centers of mass of the two chains parallel to the cylindrical
compartment as a function of time. Initially, $R_\text{c2c}$ is
zero due to the preparation of the system, and stays close to zero
during the induction time.
Eventually, segregation sets in, and $R_\text{c2c}$ grows rapidly
until $R_\text{c2c}=L_\text{eq}$ is reached, at which time the
chains do not overlap anymore. Further increase in $R_\text{c2c}$
is only due to diffusion and is therefore much slower.
\begin{figure}[tb]
  \centering
  \includegraphics[width=.45\textwidth]{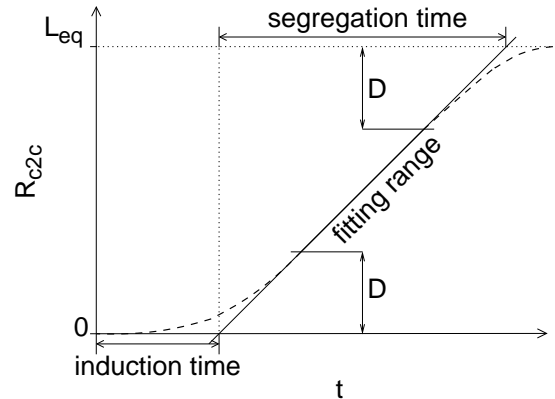}
  \caption{Schematic view of the segregation process. A line is fitted to the
    center of mass distance $R_\text{c2c}$ in the range from $D$ to $L_\text{eq}-D$. From
    this fit, the induction time $t_i$ is determined as intersection with $R_\text{c2c}=0$,
    and the segregation speed $V_\text{c2c}$ as its slope. The segregation time
    is then $t_s=L_\text{eq}/V_\text{c2c}$.}
  \label{fig:segscheme}
\end{figure}

Fig.~\ref{fig:segscheme} displays schematically how we extract the
induction and segregation times from each run: we fit a linear
function $(t - t_i)V_\text{c2c}$ to the range in which
$R_\text{c2c}(t)$ is between $D$ and $L_\text{eq}-D$. Here, $t_i$
is the extrapolated onset time of segregation and $V_\text{c2c}$
is the speed with which the two centers of mass separate in the
linear regime. We always find linear segregation behavior  for
$D\le R_\text{c2c}\le L_\text{eq}-D$. The lower limit implies that
the chains are separated by at least one blob diameter, the upper
limit guarantees that there is at least one blob-size overlap
left.

\section{Results and Discussion}

\begin{figure}[bt]
  \centering
  \includegraphics[width=.48\textwidth]{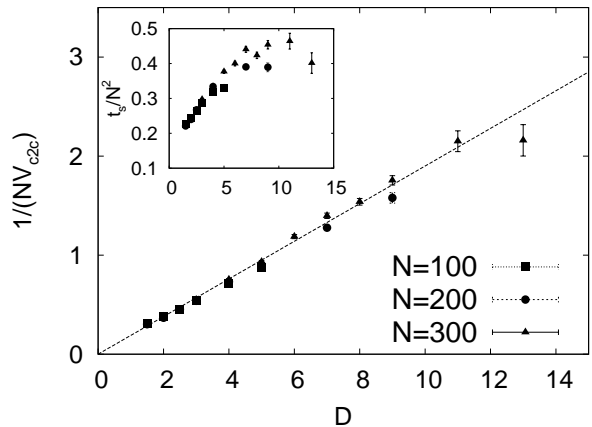}
  \caption{Measured segregation speed $V_\text{c2c}$ rescaled by $1/N$. For better
    visualization, we actually plot its inverse, $1/(N V_\text{c2c})$. The dashed
    line demonstrates the linear scaling of the segregation time with $D$. The
    inset shows the segregation time $t_s$ rescaled by $N^2$, as a function of $D$.}
  \label{fig:segtime}
\end{figure}

As can be seen in Fig.~\ref{fig:segtime}, our simulation clearly
support the scaling prediction $V_\text{c2c}\sim 1/(ND)$
(Eqn.~\eqref{eq:sseg}). The prediction that the segregation time,
scales as $N^2$ is only recovered for small tube  diameters. This
is not unexpected, because when $D$ approaches $L_\text{eq}$, the
simple blob prediction for $L_\text{eq}$ breaks
down~\cite{future} and the
segregation time levels off at the relaxation of a free chain.
The segregation speed relation Eqn.~\eqref{eq:sseg} however seems
to be quite robust even for finite systems.

\begin{figure}[tp]
  \centering
  \includegraphics[width=.48\textwidth]{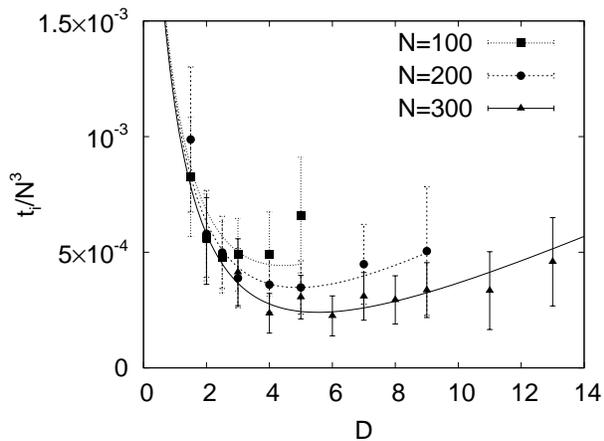}
  \caption{Measured induction times $t_i$ rescaled by $N^3$. The lines are just
    guides to the eye.}
  \label{fig:conftime}
\end{figure}

The measured average induction time $t_i$ is shown in
Fig.~\ref{fig:conftime}. The distribution of the induction times
has a long tail, which makes it difficult to sample $t_i$
accurately. Keeping this caveat in mind,  we find that for small
tube diameters the $t_i$'s computed for different chain lengths
can be made to collapse if we assume $N^3$-scaling, as expected
for a diffusive process. Moreover, we do observe the expected
decrease of  $t_i$ with increasing $D$. For larger tube diameters,
the induction time increases again; this is probably  due to the
fact that for larger $D$ the segregation and induction times
cannot be clearly separated ($t_i/t_s= {\mathcal O}(1))$ for
larger diameters. In fact, in our simulations, the induction time
seems to converge to about one quarter of the segregation time for
all $N$ and $D\gtrapprox 4$.

We stress that for highly confined chains, the diffusive process
is only responsible for the segregation over the tiny initial
separation necessary to obtain a significant effective entropic
force $F_\text{eff}$. The overwhelming part of the chain
``demixing'' is due to directed segregation (see
Fig.~\ref{fig:segrun}). In other words: it may take a while for
the system to \emph{start} segregating, but the segregation
process itself is always governed by the effective entropic force.

\begin{figure}[tp]
  \centering
  \includegraphics[width=.48\textwidth]{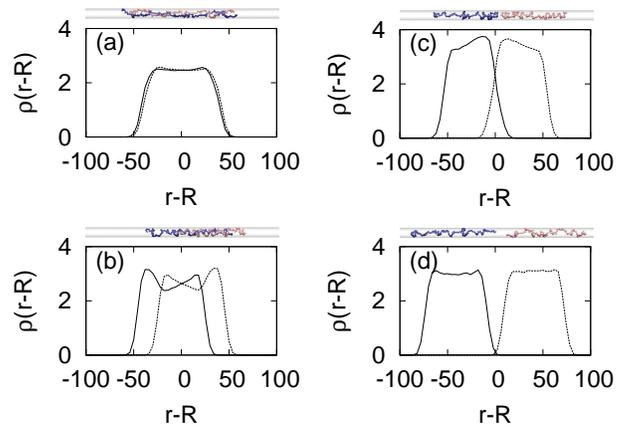}
  \caption{Monomer density $\rho(r)$ along the cylinder--axis for $N=200$,
    $D=8$, averaged over all configurations with a center of mass distance
    $R_\text{c2c}=0$ (a), $R_\text{c2c}=20$ (b), $R_\text{c2c}=48$ (c) and
    $R_\text{c2c}=80$ (d). The graphs are centered around the systems center of
    mass $R$.}
  \label{fig:densityseries}
\end{figure}

Fig.~\ref{fig:densityseries}  shows the average monomer densities
of the two polymers for different center-of-mass distances $R_\text{c2c}$.
As predicted, the monomer densities of each chain in the overlap 
region are almost unaffected by the presence of a second polymer
in the same space. Hence, the initial monomer density is almost
twice as large as for a single chain.  During segregation, the
monomer densities of the individual polymers increase somewhat. In
fact, the snapshots show that the polymers have  very nearly
separated at $R_\text{c2c}=48$, which is significantly less than
$L_\text{eq}=61.1$; this demonstrates that the polymers deform
during segregation: the entropic driving force  is strong enough
to compress the polymers. After demixing, the chains expand to
their equilibrium length.

\section{Conclusions}

Our simulations support the scaling prediction that the
entropically driven segregation of two confined chains requires a
time proportional to $N^2$. For long chains, this time is much
shorter than the diffusive time that scales as $N^3$. We stress
that this speed up of entropically driven segregation does not
involve any active (energy-consuming) process. Considering the
geometry of confinement and the length scales of (filamenteous)
bacteria, our results strongly suggest that the partitioning of
duplicated chromosomes in these organisms is, at least partly,
entropy-driven. Since the segregation sets already in during
replication, there is no initial ``induction'' regime for bacteria.
Indeed, the recent data obtained by Hu \emph{et al.} on the cyanobacterium
\emph{Anabaena} sp. PCC 7120 suggests that MreB, a bacterial actin homologue that
is speculated by some as a ``track'' for transporting chromosome by putative
motor proteins, is important for cell shape but not for chromosome
segregation~\cite{Zhao}. Moreover, they also have shown that the ratios of DNA
content in two daughter cells have a much wider distribution than in the case
that the two cells were identical. This suggests that chromosome partitioning
is a random process, in good agreement with our entropy-driven segregation
process we presented in this article.

\section*{Acknowledgments}

We thank Daan Frenkel and Bae-Yeun Ha for many helpful comments and discussions.
In addition, we are grateful to Bae-Yeun Ha for providing us computational
resources that made the simulations performed in this work possible. This work
is part of the research program of the Stichting voor Fundamenteel Onderzoek der
Materie (FOM), which is supported by the Nederlandse Organisatie voor
Wetenschappelijk Onderzoek (NWO). AA acknowledges support from the Marie-Curie
program under the European Community's Sixth framework programme, and SJ the
post-doctoral fellowships from NSERC (Canada) and Marie-Curie (EU).


\end{document}